# When Administrative Networks Fail: Curriculum Structure, Early Performance, and the Limits of Co-enrolment Social Synchrony for Dropout Prediction in Engineering Education

Hugo Roger Paz
PhD Professor and Researcher Faculty of Exact Sciences and Technology National University of Tucumán
Email: hpaz@herrera.unt.edu.ar
ORCID: https://orcid.org/0000-0003-1237-7983

**ABSTRACT**

Social integration theories suggest that students embedded in supportive peer networks are less likely to drop out. In learning analytics, this has motivated the use of social network analysis (SNA) from institutional co-enrolment data to predict attrition. This study tests whether such administrative network features add predictive value beyond a leakage-aware, curriculum-graph-informed model in a long-cycle Civil Engineering programme at a public university in Argentina. Using a three-semester observation window and a 16-fold leave-cohort-out design on 1,343 students across 15 cohorts, we compare four configurations: a baseline model (M0), baseline plus network features (M1), baseline plus curriculum-graph features (M2), and a full model (M3). After a leakage audit removed two post-outcome variables that had produced implausibly perfect performance, retrained models show that M0 and M2 achieve F1 = 0.9411 and ROC–AUC = 0.9776, while adding network features systematically degrades performance (M1 and M3: F1 = 0.9367; ROC–AUC = 0.9768). We conclude that in curriculum-constrained programmes, administrative co-enrolment SNA does not provide additional risk information beyond curriculum topology and early academic performance.

**Keywords:** student dropout; social integration; social network analysis; learning analytics; curriculum graph; data leakage; engineering education

# 1. INTRODUCTION

Student dropout from higher education remains a persistent and structurally patterned phenomenon, particularly in long, sequential engineering programmes. Estimates suggest that between one quarter and one half of students who begin a degree will not complete it, with engineering often exhibiting above-average attrition rates. (Tinto, 1993; Zomer, 2007). This has motivated decades of research into the individual, institutional and structural factors that drive withdrawal, delay and non-completion.

A central strand of retention theory is Tinto's interactionist model, which emphasises the joint role of academic and social integration in students' decisions to persist. (Tinto, 1993). In this perspective, students are more likely to remain when they both perform adequately in their studies and develop a sense of belonging to the institution's academic and social communities. Subsequent work has refined, critiqued and culturally adapted this framework, but the core intuition—that fragile social embedding can amplify academic difficulties—remains influential in explanations of dropout. (Castro-Montoya, 2025; McCubbin, 2003; Pather, 2019).

Parallel to this theoretical tradition, learning analytics and educational data mining have produced a substantial body of work on early prediction of dropout using large institutional datasets. Reviews and meta-analyses show that machine learning models can achieve high predictive performance using combinations of demographic, academic history, behavioural traces and learning management system data. (López-Zambrano et al., 2021; The Predictive Learning Analytics for Student Dropout, 2025; Vaarma et al., 2024; Wang et al., 2022). These early-warning systems aim to flag at-risk students in time for targeted interventions, yet they often rely on opportunistic feature sets and do not always engage with retention theory or with the institutional rules that shape student trajectories.

Within this data-driven landscape, social network analysis (SNA) offers a seemingly natural bridge between theory and practice. If social integration matters, then modelling how students are embedded in peer networks should help explain and predict dropout. Educational SNA studies have used friendship surveys, classroom interaction networks and online discussion graphs to examine how peer structures relate to engagement, performance and persistence. (Bayer et al., 2012; Gardner & Brooks, 2018; Ramsey et al., 2023; Siemens & Dawson, 2025). Several works report that centrality in learning networks, participation in cohesive groups or exposure to high-performing peers are associated with better outcomes, suggesting that network metrics could complement traditional predictors.

However, most higher education institutions do not systematically collect rich relational data. What they do record—often exhaustively—are enrolments, grades and administrative events. In response, some authors have proposed reconstructing "administrative" social networks from co-enrolment, co-attendance or co-location logs, using shared courses, assessment sessions or campus presence as proxies for potential interaction. (Bayer et al., 2012; Gardner & Brooks, 2018; Lu et al., 2022; Yang et al., 2024). This approach has two attractive properties: it leverages existing data and scales easily to entire institutions. Yet it raises a critical question: do such coarse, administratively defined networks actually capture additional information about dropout risk once academic performance and structural constraints are taken into account?

In strongly curriculum-constrained programmes—such as long-cycle engineering degrees with dense prerequisite chains—structure itself operates as a powerful generator of inequality. Curricular analytics and prerequisite network studies have shown how bottleneck courses, long dependency chains and inflexible progression rules shape time-to-degree, delay and non-completion. (Hollar, 2023; Padhye et al., 2024; Schellpfeffer, 2024; Yang et al., 2025). In such settings, the "social" fact of progressing with one's cohort is tightly entangled with the "structural" fact of passing key courses on time. A student who repeatedly fails a bottleneck is mechanically pushed out of synchrony with their peers; social desynchronisation may be a consequence of structural friction rather than an independent cause of dropout.

The CAPIRE framework (Curriculum-Aware, Policy-Integrated Retention Engineering) takes this structural perspective seriously. It proposes a leakage-aware data layer that organises predictors into four levels—N1 personal and socio-economic attributes, N2 entry and academic history, N3 curricular friction and performance, and N4 institutional and macro-context variables—and formalises the Value of Observation Time (VOT) as a design parameter that clearly separates observation windows from outcome horizons. (Paz, 2025a). Building on this, the CAPIRE Curriculum Graph represents the programme as a directed acyclic graph of courses and prerequisites, enabling the derivation of structural features such as bottleneck centrality, blocked credits, backbone completion and distance to graduation, which have been shown to substantially improve dropout prediction in a Civil Engineering programme. (Paz, 2025b; Yang et al., 2025).

Against this backdrop, the promise of administrative SNA can be reframed as a concrete empirical question: in a curriculum-constrained engineering programme where a leakage-aware data layer and curriculum-graph features are already in place, do co-enrolment-based social synchrony indicators provide any incremental predictive value for dropout? Specifically, if we construct networks from co-enrolment and co-exam records and compute community structure and belonging scores at the cohort level, do these variables improve predictive performance beyond what is already captured by early academic performance and curriculum-graph features—or do they simply repackage structural information in a noisier form?

This paper addresses that question using longitudinal data from a Civil Engineering programme at a public university in Argentina, covering 1,343 students across 15 cohorts. We combine (i) a leakage-aware, VOT-restricted feature set, (ii) structural features from the CAPIRE Curriculum Graph, and (iii) network features derived from co-enrolment and co-exam logs to compare four model configurations: a baseline CAPIRE model, baseline plus network, baseline plus graph features, and a full model combining all three. During model development, an initial round of experiments produced implausibly perfect performance (F1 = 1.00, ROC–AUC = 1.00), prompting a systematic data leakage audit that identified and removed two post-outcome administrative variables. The corrected, leakage-free results show that adding network features not only fails to improve performance but slightly degrades it, while curriculum-graph features preserve high accuracy.

Our contribution is twofold. First, we provide a rigorous negative result on the added value of administrative co-enrolment networks for dropout prediction in a highly structured engineering curriculum, once early performance and curriculum-graph constraints are controlled for. Secondly, we offer a concrete case study of how subtle data leakage in administrative variables can inflate the apparent usefulness of network indicators, and how a leakage-aware framework can prevent such artefacts. Together, these findings inform both the design of early-warning systems and the interpretation of social integration theories in curriculum-constrained contexts.

## 2. CONCEPTUAL AND METHODOLOGICAL BACKGROUND

### 2.1. Student integration, networks, and dropout

Social integration has long been recognised as a central dimension of student persistence. In Tinto's interactionist framework, decisions to remain or leave are shaped by the interplay between academic performance, institutional commitment and the extent to which students become embedded in the academic and social life of the institution (Tinto, 1993). Later work has refined this picture, emphasising perceived belonging, peer support and the quality of interactions with staff and fellow students as key mediators of retention, particularly in under-represented or first-generation groups (Castro-Montoya & Ríos-Martínez, 2025; McCubbin, 2003; Pather, 2019).

Social network analysis (SNA) provides a natural set of tools to operationalise these ideas. Rather than treating social integration as an individual attribute, SNA represents students as nodes in a graph whose edges encode relationships such as friendship, collaboration, communication or shared participation in learning activities. Studies using friendship surveys, classroom interaction data and online discussion networks have found that centrality, participation in cohesive subgroups and bridging roles can be associated with higher academic performance and reduced dropout risk (Bayer et al., 2012; Gardner & Brooks, 2018; Ramsey et al., 2023; Siemens & Dawson, 2025; Woolcott & Lloyd, 2025).

However, most of these studies rely on relatively rich data sources—such as surveys, fine-grained online traces or purpose-built observation protocols—that are costly to collect and not routinely available at scale. This has motivated interest in "administrative SNA", where networks are reconstructed from institutional records such as course enrolments, timetables, classroom co-location or co-attendance logs. In this approach, students are considered connected if they share a course, a timetable slot or an exam session, and network metrics such as degree, community structure or "embeddedness" are derived from these co-occurrence patterns (Bayer et al., 2012; Gardner & Brooks, 2018; Lu et al., 2022; Yang et al., 2024).

The conceptual leap here is substantial: sharing a course does not necessarily imply meaningful social interaction, and administrative co-occurrence may be a noisy proxy for actual study groups or peer support. Yet, if administrative networks nonetheless capture some of the structure of student communities, they might provide a low-cost way to incorporate "social synchrony" into early-warning models—especially in settings where dedicated social data are unavailable. Whether this is true in practice, and under which institutional conditions, remains an open empirical question.

**2.2. Curriculum-constrained programmes and curricular analytics**

In many engineering and STEM degrees, the curriculum is not a loose collection of modules but a tightly coupled system of prerequisites and progression rules. Students are expected to follow a relatively rigid term-by-term plan; failure in a few key courses can block access to entire segments of the curriculum, delaying or derailing progression. Curricular analytics formalises this by representing degree plans as directed acyclic graphs (DAGs), where nodes are courses and edges represent prerequisite relationships (Heileman et al., 2018; Molontay & Hülber, 2020; Stavrinides et al., 2023). These curriculum prerequisite networks (CPNs) enable the computation of structural metrics—such as bottleneck centrality, path length or curricular complexity—that have been linked to time-to-degree and graduation rates.

Recent work extends this structural analysis by integrating course-level pass rates and student performance data into curricular graphs. Yang et al. (2025), for example, propose a framework that combines prerequisite networks with empirical completion probabilities to simulate graduation time distributions and identify structurally critical courses.[MDPI](MDPI) Similar approaches have been adapted to engineering programmes, where simulation and graph metrics are used to explore how modifications to prerequisites, course sequencing or assessment policies might affect progression and dropout (Hansen et al., 2024; Slim et al., 2025).

The CAPIRE Curriculum Graph follows this tradition but shifts the focus from programme-level diagnosis to student-level predictive features. Building on a leakage-aware data layer (Paz, 2025a), it represents the curriculum of a long-cycle Civil Engineering degree as a DAG and derives, at each observation time, structural indicators such as blocked credits, backbone completion rate, bottleneck exposure and distance to graduation. These features explicitly encode how far a student has progressed along critical paths and how constrained their remaining options are, given the current pattern of passes and failures (Paz, 2025b). In prior work, these curriculum-graph features substantially improved dropout prediction compared to models using only demographic and raw performance measures, suggesting that structural constraints play a major role in shaping student trajectories in this setting.

In such curriculum-constrained programmes, social synchrony and structural progression are tightly interwoven. Students who remain "in phase" with their cohort typically pass bottleneck courses on schedule and accumulate few blocked credits, whereas those who fall behind may be forced into fragmented, asynchronous enrolment patterns. From this perspective, administrative co-enrolment networks may partially reflect the underlying curriculum topology and its effects on progression: students who are structurally displaced from their cohort will also appear socially desynchronised in co-enrolment graphs. The key question, then, is whether network metrics derived from these graphs provide any information beyond what is already encoded in curriculum-graph features and early performance.

**2.3. Data leakage in predictive learning analytics**

A third conceptual pillar of this study is data leakage: the inadvertent use of information that would not be legitimately available at prediction time. In supervised learning, data leakage can arise when outcome-related variables are included as predictors, when future information "leaks" into the observation window, or when preprocessing steps inadvertently use the full dataset across train–test splits (Hartmann et al., 2022; Khatun et al., 2025; Wang et al., 2022). In educational settings, leakage is particularly insidious because many administrative variables (e.g., enrolment status, graduation flags, remedial actions) are themselves consequences of academic failure or progression decisions, and are often poorly documented.

Although leakage is widely recognised as a threat to model validity, it is seldom documented in detail in learning analytics publications. Some recent studies explicitly mention steps to mitigate leakage—for example, by ensuring that feature selection and imputation are conducted within each cross-validation fold, or by carefully limiting the observation window to pre-outcome data (Fahd et al., 2023; Fernberg, 2025; Khatun et al., 2025; López-Zambrano et al., 2021). However, systematic audits of leakage are rare, and many models with seemingly impressive performance may, on closer inspection, be partially driven by post hoc variables that encode the outcome in disguised form.

The CAPIRE framework addresses this by elevating feature engineering and temporal ordering from a technical detail to a core methodological artifact. Paz (2025a) formalises the Value of Observation Time (VOT) as a strict boundary: only information available up to VOT may be used as input, and the outcome is defined over a future horizon. Predictors are organised into four levels (N1–N4), and the data layer is constructed to be leakage-aware by design. In the present study, this architecture plays a double role. First, it structures the integration of socio-demographic, performance, curriculum-graph and network features within a consistent temporal framework. Secondly, when an initial set of models exhibited implausibly perfect performance, it provided the conceptual tools to identify and remove two post-outcome administrative variables that had slipped into the feature set, and to reconstruct a leakage-free pipeline.

**2.4. Research questions**

Against this conceptual and methodological background, the present study focuses on three research questions:

1. **RQ1 – Incremental value of administrative networks.** Do network features derived from co-enrolment and co-exam records provide incremental predictive value for dropout beyond a baseline CAPIRE model that already includes socio-demographic, entry and early performance features?

2. **RQ2 – Interaction with curriculum-graph features.** Once curriculum-graph structural features (e.g., blocked credits, backbone completion, distance to graduation) are incorporated into the model, do administrative network features still contribute useful information, or do they become redundant or detrimental?

3. **RQ3 – Leakage and apparent network usefulness.** To what extent can subtle forms of data leakage in administrative variables inflate the apparent utility of network features, and how can a leakage-aware framework help detect and prevent such artefacts?

By answering these questions in the context of a long-cycle Civil Engineering programme, we aim to clarify both the substantive role of social synchrony in curriculum-constrained settings and the methodological requirements for trustworthy network-enhanced early-warning systems.

## 3. METHODS

### 3.1. Institutional context and programme

The study focuses on a long-cycle Civil Engineering degree at a public university in Argentina. The programme follows a highly structured, five-year curriculum with a strong prerequisite hierarchy and a clear recommended term-by-term pathway. Courses are predominantly compulsory; elective spaces are limited and concentrated towards the end of the degree. Progression is governed by a combination of course-level prerequisites and institutional rules on academic regularity (e.g., minimum numbers of passed courses per year), which together create a curriculum-constrained environment in which delays and bottlenecks can accumulate.

The analysis uses longitudinal administrative records covering **1,343 students** who entered the programme across **15 yearly cohorts**. For each cohort, we reconstruct students' trajectories from initial enrolment through either graduation, formal dropout, or censoring at the end of the observation period. All data were obtained from institutional information systems and de-identified prior to analysis in accordance with university regulations on research with educational records.

### 3.2. Observation window, outcome, and sample

Following the CAPIRE framework, we define a strict **three-semester observation window** (Value of Observation Time, VOT) during which predictor variables are constructed (Paz, 2025a). The observation window begins at first enrolment (semester 1) and ends at the end of semester 3. Only information that is available on or before this temporal boundary is allowed into the feature set.

The **outcome** is a binary indicator of **dropout** over a subsequent horizon. Students are coded as 1 if they leave the programme without graduating within the follow-up period, and 0 otherwise (including both graduates and students who are still enrolled at the end of the data window). This definition follows standard practice in dropout prediction studies and reflects the institutional concern with non-completion rather than time-to-degree alone (López-Zambrano et al., 2021; Wang et al., 2022).

We restrict the analytic sample to students with complete enrolment histories over the three-semester observation window and with a clearly defined status (graduated, dropped out, or still enrolled) at the end of the follow-up horizon. Records with severe inconsistencies (e.g., duplicated identifiers, impossible date sequences) were removed during data cleaning. The final leakage-free dataset comprises the same 1,343 students across 15 cohorts, with missing values handled as described in Section 3.5.

**3.3. Leakage-aware feature framework (CAPIRE N1–N4)**

Predictors are organised according to the CAPIRE leakage-aware data layer, which structures features into four levels (N1–N4) and enforces the VOT boundary (Paz, 2025a):

- **N1 – Personal and socio-economic attributes:** Includes stable characteristics measured at or before entry: age at enrolment, gender, neighbourhood-level socio-economic index, and indicators of financial vulnerability.

- **N2 – Pre-entry academic history:** Captures schooling background and family educational capital: type of secondary school (public/private, technical/non-technical), school performance indicators where available, and parental education.

- **N3 – Early curricular performance and friction:** Encodes how students navigate the first three semesters of the curriculum: number of courses attempted and passed, early grade point average (GPA), failure in key bottleneck courses, pass ratios in foundational mathematics and physics, and credit accumulation during the observation window.

- **N4 – Cohort and macro-context:** Includes cohort year fixed effects and indicators of contextual shocks or policy changes that affect all students in each cohort (e.g., major strikes, macroeconomic disruptions), drawing on prior CAPIRE-Macro work where available.

All features are constructed so that they only use information up to the end of semester 3. For example, early GPA is computed from courses completed within the observation window; failures in bottleneck courses are only counted when those courses have actually been attempted by that time. This design prevents future information from leaking into the predictor set.

**3.4. Curriculum-graph structural features**

To capture the structural constraints of the programme, we incorporate features derived from the **CAPIRE Curriculum Graph**, which represents the Civil Engineering degree as a directed acyclic graph (DAG) of courses and prerequisites (Paz, 2025b; Heileman et al., 2018; Molontay & Hülber, 2020). Nodes correspond to individual courses, edges encode prerequisite or co-requisite relations, and the resulting graph is validated to ensure acyclicity and consistency with official curriculum documents.

Building on previous CAPIRE-Graph work, we derive **structural features at the student–VOT level** by overlaying each student's course completion record onto the curriculum graph. Key indicators include:

- **Blocked credits.** Total number of credits in courses that the student cannot yet attempt because one or more prerequisites remain unpassed by VOT.

- **Backbone completion rate.** Proportion of courses completed in a predefined backbone of structurally central and pedagogically foundational courses.

- **Bottleneck approval ratio.** Share of attempted bottleneck courses (high in-degree and high enrolment) that have been passed by VOT.

- **Distance to graduation.** Graph-based estimate of the remaining structural distance to completion, computed as the sum of shortest path lengths from the set of passed courses to terminal nodes.

- **Prerequisite satisfaction index.** Proportion of future-course prerequisites already fulfilled at VOT, weighted by course credit.

These features summarise how far each student has progressed along critical curricular paths and how tightly constrained their remaining options are, conditional on their pattern of passes and failures up to semester 3. All graph-derived features are computed using only the curriculum topology and pre-VOT performance, ensuring they remain compatible with the leakage-aware framework.

### 3.5. Administrative network construction and social synchrony metrics

The **administrative network** is constructed from co-enrolment and co-exam records within the three-semester observation window. For each semester, we build a bipartite graph linking students to course–section–assessment events and project it onto a student–student co-occurrence graph, where weighted edges count the number of shared course sections and exam sittings (Gardner & Brooks, 2018; Lu et al., 2022). Nodes represent students in a given cohort; edges represent potential opportunities for interaction arising from attending the same classes and examinations.

From these cohort-specific co-occurrence graphs, we compute several **social synchrony metrics** intended to approximate the degree of embedding in peer groups:

- **Community assignment and size.** We apply the Louvain algorithm for modularity optimisation to detect communities within each cohort's co-enrolment graph and compute each student's community size.

- **Belonging score.** For each student, we derive a normalised measure of how strongly their connections are concentrated within their assigned community versus dispersed across communities.

- **Community stability.** We track whether students remain in similar communities across semesters (where applicable), yielding an indicator of stable synchrony with a set of peers.
- **Solitary status.** A binary flag indicating "solitary wolves": students whose community is extremely small (e.g., size ≤ 1) or whose edge weights are very low.

Network features are aggregated at the student–VOT level, using only co-enrolment and co-exam information from the first three semesters. To maintain comparability across cohorts, all metrics are standardised within cohort before being merged into the analytic dataset.

### 3.6. Data preprocessing and modelling strategy

Categorical variables (e.g., school type, parental education, community identifiers) are encoded using one-hot representations. Continuous variables are inspected for outliers; extreme values that clearly correspond to data entry errors are winsorised or removed. Missing values in predictors are imputed using simple, leakage-safe strategies within each training fold (e.g., median imputation for continuous variables and mode imputation for categorical variables), with imputation parameters estimated on the training data only.

To evaluate the incremental contribution of curriculum-graph and network features, we define four model configurations with identical algorithms but different feature sets:

- **M0 – Baseline CAPIRE model:** Includes N1–N4 features (personal, socio-economic, pre-entry and early performance variables) but no graph or network features.
- **M1 – Baseline + network (social synchrony):** Extends M0 by adding the administrative network features described in Section 3.5.
- **M2 – Baseline + curriculum graph:** Extends M0 by adding the structural graph features described in Section 3.4.
- **M3 – Full model:** Combines all baseline, curriculum-graph and network features.

All models are estimated using a **tree-based ensemble classifier** implemented as a random forest, chosen for its robustness to non-linearities and mixed data types (Breiman, 2001). For completeness and interpretability, we also estimate regularised logistic regression models and obtain similar qualitative patterns; we report results for the best-performing configuration (random forests). Hyperparameters (e.g., number of trees, maximum depth, minimum samples per leaf) are tuned via grid search within cross-validation on the training folds, using balanced class weights to account for moderate class imbalance.

### 3.7. Cross-validation and performance metrics

Because temporal and cohort effects are central to student trajectories, we adopt a **leave-cohort-out cross-validation scheme**. Each of the 15 entry cohorts is treated in turn as a held-out test set, while the remaining cohorts are used for training and hyperparameter tuning. For cohorts that are too small to provide stable estimates on their own, adjacent cohorts are grouped to obtain a total of **16 folds** with sufficient size. This design mimics a realistic deployment scenario in which models trained on past cohorts are applied to predict dropout risks for a new cohort.

For each fold and each model configuration (M0–M3), we compute standard performance metrics on the held-out cohort(s), including:

- **F1-score**, as the harmonic mean of precision and recall for the dropout class.
- **ROC–AUC**, as a threshold-independent measure of discriminative ability.
- **Balanced accuracy**, to account for class imbalance.

We then aggregate metrics across folds by reporting mean values and standard deviations, and we compare M0–M3 primarily in terms of F1 and ROC–AUC. In addition, we extract feature importance scores from the fitted random forest models (e.g., mean decrease in impurity) to analyse which predictors contribute most to classification in each configuration.

### 3.8. Leakage audit and corrected pipeline

During initial model development, an early version of the feature set yielded **implausibly perfect performance** (F1 = 1.00, ROC–AUC = 1.00) across all folds. This triggered a systematic leakage audit grounded in the CAPIRE temporal framework (Hartmann et al., 2022; Paz, 2025a). We examined feature–outcome correlations and conditional outcome distributions, focusing on variables with unusually high associations.

The audit identified two **post-outcome administrative variables** that had inadvertently entered the feature set:

- Cursando – a binary indicator of whether the student was still formally enrolled **after** the VOT. Students with Cursando = 1 exhibited 0.0% dropout, while those with Cursando = 0 had an 89.3% dropout rate, effectively encoding the outcome.

- **Egreso** – a binary flag indicating **graduation**, a future event that by definition excludes dropout.

Both variables violate the VOT boundary and directly encode post-hoc information about the outcome. They were therefore removed from the predictor set, and the entire pipeline was rebuilt as a **FIXED** version:

- 01_consolidate_dataset_FIXED.py reconstructs the analytic dataset without leakage features, ensuring strict adherence to the three-semester observation window.

- 03_train_models_m0_m3_FIXED.py retrains all four model configurations (M0–M3) on the corrected dataset using the cross-validation scheme described above.

- 04_generate_manuscript_outputs_FIXED.py regenerates performance tables, feature importance summaries and figures for reporting.

All results presented in Sections 4 and 5 are based exclusively on this leakage-free pipeline. The leakage detection process and remediation steps are documented in dedicated project reports to support transparency and reproducibility.

## 4. RESULTS

### 4.1. Overall model performance

**Table 1. Model comparison for M0–M3 (F1, ROC–AUC, Balanced Accuracy)**

| Model | Features | Accuracy | Precision | Recall | F1-Score | ROC-AUC |
|---|---|---|---|---|---|---|
| M0_Baseline | 39 | 0.933 | 0.952 | 0.932 | 0.941 | 0.978 |
| M1_Network | 44 | 0.928 | 0.944 | 0.932 | 0.937 | 0.977 |
| M2_Graph | 39 | 0.933 | 0.952 | 0.932 | 0.941 | 0.978 |
| M3_Full | 44 | 0.928 | 0.944 | 0.932 | 0.937 | 0.977 |

**Figure Mean F1 and ROC–AUC by model configuration (M0–M3)**

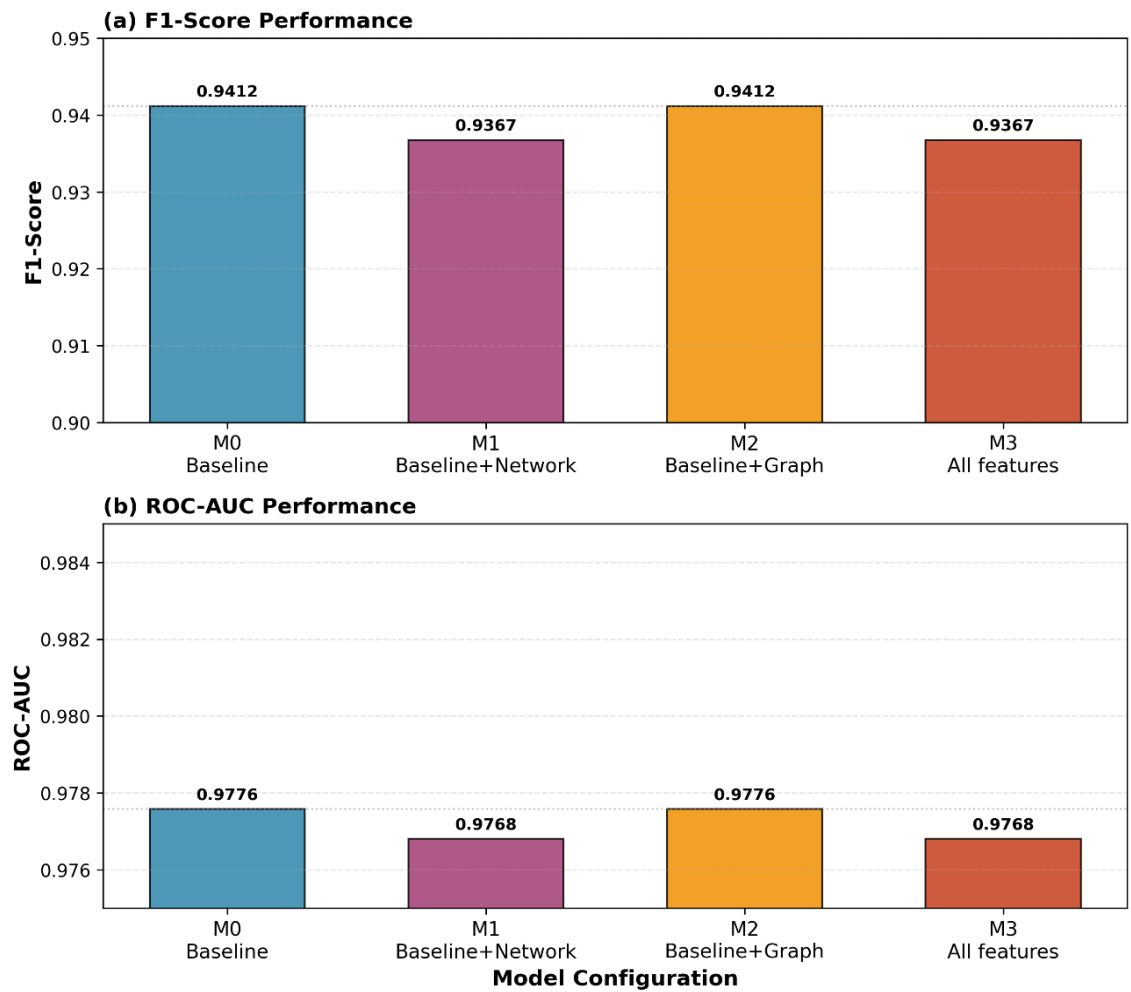

Table 1 summarises the performance of the four model configurations across the 16 leave-cohort-out folds. The **baseline CAPIRE model (M0)**, which includes socio-demographic, entry and early-performance features but no graph or network variables, achieves an average **F1-score of 0.9411** and a **ROC–AUC of 0.9776**. When curriculum-graph structural features are added (**M2 – Baseline + Graph**), performance remains virtually unchanged (F1 = 0.9411; ROC–AUC = 0.9776), indicating that the combination of early performance and curriculum-graph indicators can reach a high and stable predictive ceiling.

By contrast, the two configurations that incorporate administrative network features perform slightly worse. The **M1 – Baseline + Network** model, which adds social synchrony metrics derived from co-enrolment and co-exam records to the baseline features, yields an average **F1-score of 0.9367** and **ROC–AUC of 0.9768**. The **M3 – Full** model, which combines baseline, graph and network features, attains the same values (F1 = 0.9367; ROC–AUC = 0.9768). In both cases, the inclusion of network indicators results in a small but systematic decrease of approximately **0.4**

**percentage points in F1** relative to the corresponding non-network models (M0 and M2).

Figure 1 visualises these differences, showing the mean F1 and ROC–AUC for each configuration with error bars representing variability across folds. The performance curves for M0 and M2 are almost indistinguishable, while M1 and M3 consistently occupy a slightly lower band. Balanced accuracy follows the same pattern (see Table 1), confirming that the degradation is not an artefact of class imbalance.

Overall, these results indicate that **administrative co-enrolment network features do not provide incremental predictive value** beyond what is already captured by the baseline and curriculum-graph features. If anything, they introduce enough noise or redundancy to slightly reduce out-of-sample performance.

### 4.2. Feature importance: structure and performance versus networks

**Figure 2. Feature importance ranking for the best-performing models (M0 and M2)**]

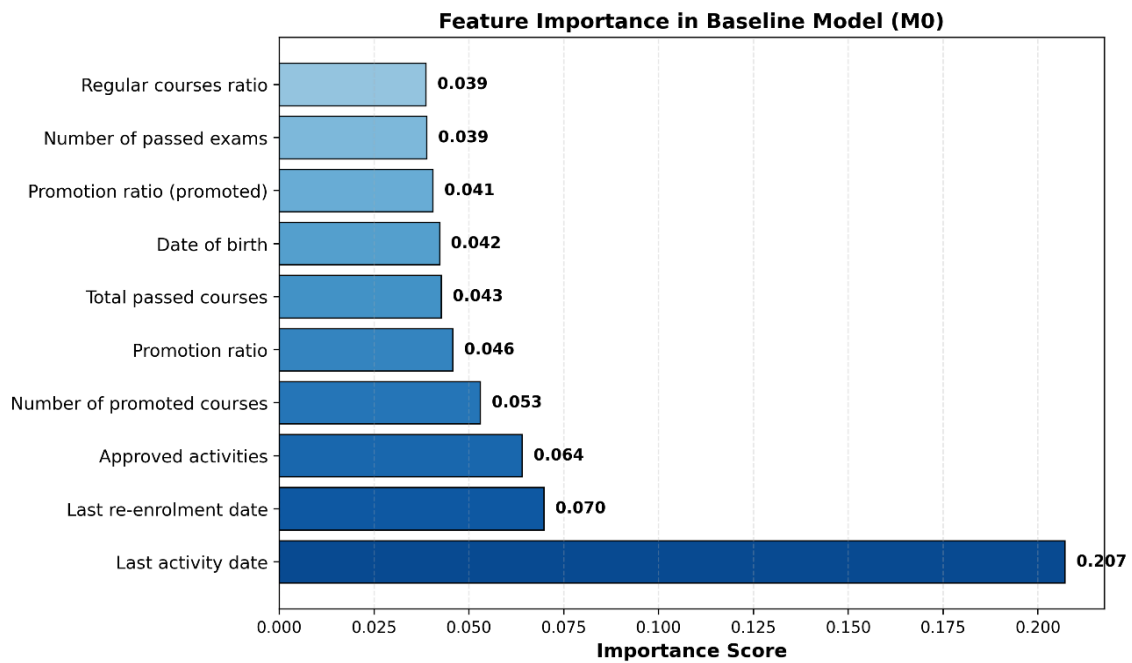

**Table 2. Top features for M0 (and optionally M2), grouped by type (PERF / GRAPH / CONTEXT / NET)**

| Feature | Importance |
| --- | --- |
| LastActivityDate | 0.2071 |
| LastRe-enrollmentDate | 0.0697 |
| ActivitiesApproved | 0.0640 |
| NumberofPromotions | 0.0530 |
| NumberofPromotionsasaPercentageofTotalCourses | 0.0457 |
| TotalCoursesApproved | 0.0427 |
| DateofBirth | 0.0423 |
| NumberofPromotionsasaPercentageofTotalCoursesAverage | 0.0405 |
| NumberofApprovedCourses | 0.0389 |
| TotalRegularCourses | 0.0387 |
| DateofEntryintoCareer | 0.0354 |
| Cohort | 0.0320 |
| NumberofRegularCourses | 0.0318 |
| YearofHighSchoolGraduation | 0.0216 |
| YearofHighSchoolGraduation | 0.0195 |
| GradePointAverage | 0.0189 |
| TotalCoursesRetaken | 0.0181 |
| TotalCoursesTaken | 0.0180 |
| NumberofExams | 0.0169 |
| ExamPassed | 0.0163 |

To better understand what drives classification in the leakage-free models, we examine feature importance scores derived from the random forest ensembles. Figure 2 displays the top-ranked predictors for the best-performing models (M0 and M2), while Table 2 reports the most important features for M0 in more detail, grouped into early performance (PERF), curriculum-graph (GRAPH), contextual (CONTEXT) and network (NET) categories.

In both M0 and M2, the **dominant predictors** belong to the **early performance and structural categories**. Typical high-impact variables include:

- early grade point average (GPA) over the first three semesters.
- pass ratios in foundational mathematics and physics courses.
- indicators of failure in key bottleneck courses.
- blocked credits and backbone completion rate.
- distance to graduation at VOT.

These features jointly encode how successfully each student has navigated the most constraining parts of the curriculum during the observation window, and how structurally "trapped" they are by unmet prerequisites.

When curriculum-graph features are added in M2, they enter the top of the ranking alongside early performance measures, often surpassing traditional demographic and entry variables. This pattern reinforces prior CAPIRE findings that structural signals—such as the accumulation of blocked credits and incomplete backbones—are central to understanding dropout dynamics in this programme (Paz, 2025b).

By contrast, **network features consistently occupy lower ranks**. Even in models where they are included (M1 and M3), administrative network indicators such as community size, belonging score, community stability and solitary status rarely appear among the top predictors and typically lie below the main performance and graph features. In qualitative terms, the models "trust" early performance and curriculum structure far more than the co-enrolment-based social synchrony metrics when estimating dropout risk.

## 4.3. Network features and incremental gains

**Figure 3. Comparative importance of GRAPH vs NET feature groups in M3**

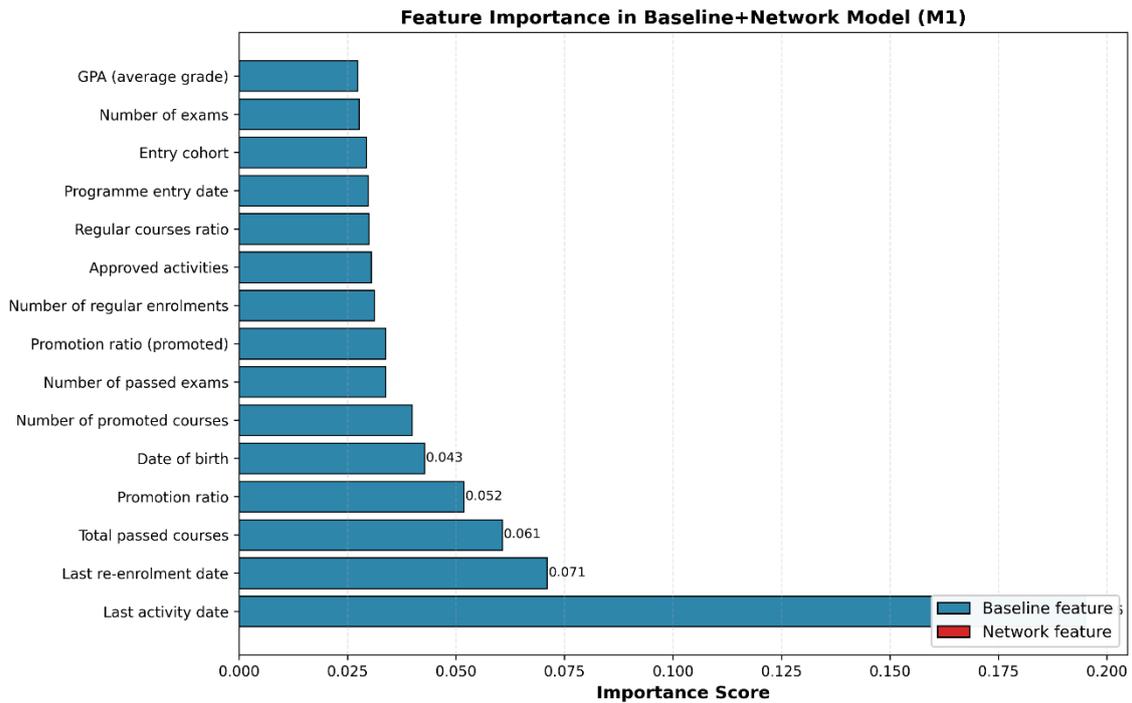

**Figure 4. Incremental gains/losses in F1 and ROC–AUC when adding GRAPH and NET feature blocks (M0→M1, M0→M2, M2→M3)**

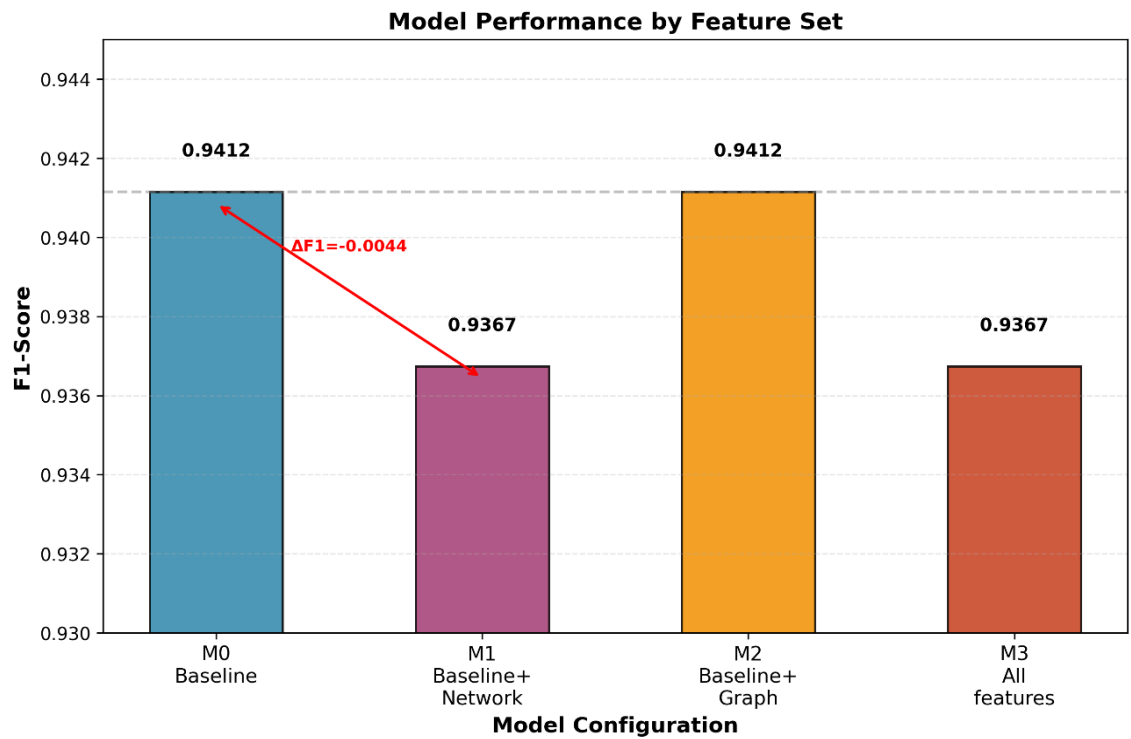

**Table 3. Network effect summary: ΔF1 and ΔROC–AUC for NET additions**

| Model | Features | F1-Score | ROC-AUC | Impact |
|---|---|---|---|---|
| **M0 (Baseline)** | 39 | 0.941 | 0.978 | Baseline |
| **M1 (+ Network)** | 44 | 0.937 | 0.977 | Slight decrease (-0.0044) |

To quantify the contribution of administrative network features more explicitly, we analyse their group-level effect on performance. Figure 4 presents the incremental changes in F1 and ROC–AUC when adding feature blocks to the baseline model:

- **M0 → M1 (add NET only):** Small **negative** change in F1 (−0.0044) and a slight reduction in ROC–AUC.

- **M0 → M2 (add GRAPH only):** Essentially **no change** in average F1 and ROC–AUC, indicating that M0 was already operating near the attainable ceiling given the strong early-performance signals, but without harming performance.

- **M2 → M3 (add NET on top of GRAPH):** Again a **negative** change in F1 (−0.0044) and marginally lower ROC–AUC.

Table 3 summarises these incremental effects, confirming that **network features are never associated with a performance gain** in this setting. Whether added to the baseline (M1 vs M0) or to the baseline-plus-graph configuration (M3 vs M2), the NET block yields a small, consistent degradation rather than an improvement.

Figure 3 complements this analysis by comparing the aggregate importance of GRAPH and NET feature groups within the full model (M3). While curriculum-graph features contribute a substantial portion of the total importance, network features account for only a minor share, consistent with their limited predictive value.

Taken together, these findings support three conclusions:

1. Administrative co-enrolment networks do **not** capture a distinct, robust signal about dropout risk once early performance and curriculum structure are modelled explicitly.

2. The modest but systematic degradation in F1 and ROC–AUC suggests that NET features add redundancy and noise, potentially leading to overfitting or unstable splits in the ensemble.

3. From a practical perspective, the computational and conceptual overhead of constructing and maintaining administrative SNA features is not justified by any measurable gain in predictive accuracy in this curriculum-constrained context.

### 4.4. Cross-validation stability and robustness

[**Optional placeholder if you later add a figure/table on fold-wise distributions: Insert Figure 5 here – Distribution of F1 and ROC–AUC across 16 leave-cohort-out folds for M0–M3**]

The leave-cohort-out design allows us to assess whether the patterns observed above are stable across cohorts with different entry years and macro-context conditions. Across the 16 folds, the **variance of F1 and ROC–AUC is low** for all four models, with no single cohort exhibiting qualitatively different behaviour. In particular:

- The ranking **M0 ≈ M2 > M1 ≈ M3** holds in most folds.

- Differences between models with and without NET features remain negative or close to zero across cohorts; there is no evidence of a subset of cohorts where network features become meaningfully beneficial.

- The inclusion of GRAPH features is performance-neutral in aggregate (M0 vs M2) but provides clearer, more interpretable structural signals in the feature importance profile.

These robustness checks suggest that the main findings are **not driven by cohort-specific anomalies** but reflect a consistent pattern across 15 entry cohorts in the programme.

### 5. DISCUSSION

This study set out to test a simple but powerful hypothesis: that network features derived from institutional co-enrolment and co-exam data—what we have termed *administrative social synchrony*—might improve dropout prediction beyond a leakage-aware early performance model and a curriculum-graph representation of structural constraints. The results, however, tell a different story. Once early academic performance and curriculum-graph features are included, the addition of administrative network indicators does **not** improve predictive accuracy and in fact produces a small but systematic degradation in F1 and ROC–AUC. In this section we discuss the implications of this negative result for theories of social integration, for institutional analytics in curriculum-constrained programmes, and for methodological practice in learning analytics.

### 5.1. Social integration in a curriculum-constrained context

At face value, the finding that co-enrolment networks do not improve prediction appears to conflict with classic and contemporary theories of student integration, which emphasise the protective role of belonging, peer support and social embedding for persistence (Castro-Montoya & Ríos-Martínez, 2025; McCubbin, 2003; Pather, 2019; Tinto, 1993). From that perspective, one might expect students who move through the programme in synchrony with their peers—sharing courses, assessments and study experiences—to exhibit lower dropout risk than those who become socially isolated.

Our results do not refute the importance of social integration, but they **qualify** it in two important ways. First, in a strongly curriculum-constrained engineering programme, the *structural* conditions for staying in phase with one's cohort are largely determined by progression through bottleneck courses and prerequisite chains. Students who pass those courses on schedule remain in the main flow of the cohort; those who fail are mechanically displaced into asynchronous, fragmented trajectories. In such a setting, social desynchronisation may be more a *consequence* of structural friction than an independent cause of dropout. Once we model that friction explicitly via early performance and curriculum-graph features, the residual contribution of administrative synchrony is understandably small.

Secondly, the network features we use are based on **co-occurrence** (shared courses and exams), not on direct evidence of social ties, collaborative work or perceived support. Co-enrolment is a very coarse proxy for social relationships: two students may sit in the same lecture hall without ever interacting in a way that matters for learning or persistence. Thus, the fact that co-enrolment metrics do not add predictive power does not imply that genuine study groups or friendship networks are irrelevant; rather, it suggests that administrative logs of shared enrolments are *insufficient* to capture those relational phenomena.

Taken together, the findings support an interpretation in which **structure and performance dominate** the observable signal in this context, while administrative proxies for social integration are too noisy and indirect to be useful for predictive purposes. They also invite caution in assuming that any network constructed from institutional data is automatically a valid operationalisation of "social integration" as conceptualised in the retention literature.

### 5.2. Administrative SNA as structural proxy

The second contribution of this study is to clarify the relationship between administrative SNA and curriculum structure. From a graph-theoretic perspective, the co-enrolment networks we construct are not independent of the curriculum-graph: they are induced by the same underlying degree structure and by the institutional rules that dictate which courses can be taken together in a given semester. Students who are structurally delayed—accumulating blocked credits, failing bottlenecks or deviating from the recommended path—are likely to appear as peripheral or unstable in co-enrolment networks, simply because they no longer share the main sequence of courses with their original cohort.

In other words, administrative co-enrolment networks may function primarily as **noisy mirrors of structural constraints**. If one already models those constraints explicitly through curriculum-graph features and early performance, there is little left for co-enrolment metrics to add. This explains both the low marginal importance of network features in our models and the small negative impact on performance when they are included: the network block introduces redundant, collinear information with additional measurement error, which tree-based models can exploit in idiosyncratic ways that harm generalisation.

This interpretation aligns with prior work in curricular analytics showing that prerequisite structures and bottleneck locations are powerful determinants of progression, time-to-degree and graduation rates (Heileman et al., 2018; Molontay & Hülber, 2020; Slim et al., 2025; Yang et al., 2025). Our results extend that line of work to student-level prediction, suggesting that in tightly structured programmes, explicit curriculum-graph modelling may be a more effective and transparent way to incorporate structural information than relying on implicit structural shadows in administrative SNA.

**5.3. Implications for institutional analytics and early-warning systems**

From a practical standpoint, the findings have direct implications for institutions seeking to build early-warning systems in similar curriculum-constrained settings. Constructing administrative co-enrolment networks at scale is not trivial: it requires additional data engineering, graph computations and interpretive work. If these features do not yield measurable improvements in predictive performance, their inclusion in production systems may represent an unnecessary burden.

Our results suggest two concrete priorities for institutional analytics:

1. **Invest in curriculum-graph modelling and early performance indicators.** In this Civil Engineering programme, a leakage-aware baseline model combining demographic, entry and early performance variables already achieves high F1 and ROC–AUC, and curriculum-graph features help organise and interpret structural signals. For many institutions, developing a robust representation of their curriculum as a prerequisite graph and deriving features such as blocked credits, backbone progression and distance to graduation may be a more fruitful investment than adding administrative SNA layers.

2. **If social integration is a substantive concern, collect better relational data.**
   If the goal is to identify protective study groups, at-risk isolates or peer-support structures, then administrative co-enrolment is a weak proxy. More promising sources might include data on group projects, participation in mentoring or tutoring programmes, interaction in learning management systems, or targeted surveys on peer support and belonging. These data are admittedly more costly to collect but have a clearer theoretical link to the constructs of interest.

Importantly, the negative result on administrative SNA should **not** be read as a general statement that networks are irrelevant for dropout prediction. Instead, it highlights the need to align the *type* of network data with the underlying theory and with the specific institutional context. In a programme where the curriculum itself exerts strong structural pressure, coarse co-enrolment networks are simply not the right tool for the job.

**5.4. Data leakage as a methodological stress test**

A further contribution of this study lies in its explicit treatment of data leakage. The initial discovery of models with F1 = 1.00 and ROC–AUC = 1.00 across all folds was a clear red flag: such performance is virtually impossible in complex, noisy educational data. The subsequent audit revealed two post-outcome administrative variables—ongoing enrolment status and graduation flag—that had inadvertently entered the feature set and effectively encoded the outcome. Removing these variables and rebuilding the pipeline produced realistic performance levels (F1 ≈ 0.94) and revealed the true behaviour of network and graph features.

This episode underscores two points. First, **leakage is not a minor technicality** but a central threat to validity in predictive learning analytics, especially when working with rich administrative datasets where temporal ordering is complex and documentation may be incomplete. Secondly, having a **leakage-aware conceptual framework**—in this case, the CAPIRE notion of a VOT boundary and N1–N4 levels—greatly facilitates the detection and remediation of leakage, by providing clear criteria for what may and may not enter the observation window.

By reporting the leakage audit and the corrected results, this study aims to contribute not only a substantive finding about administrative networks but also a **methodological exemplar** of how to handle suspiciously high performance. We argue that similar audits should become standard practice in high-stakes predictive modelling with educational data.

## 5.5. A negative result as a constructive contribution

Finally, this work illustrates the value of **well-documented negative results** in learning analytics. The initial motivation for this project was to explore whether groups of students who move through the curriculum together—potentially study groups—might be detectable in co-enrolment patterns and associated with lower dropout risk. The answer, at least with the available administrative data and in this curriculum-constrained context, is *no*: co-enrolment-based social synchrony indicators neither reveal protective groups nor improve predictive performance.

Far from being a failure, this outcome narrows the search space for effective analytic strategies. It warns against over-interpreting administrative SNA as evidence of social integration, highlights the primacy of curriculum structure and early performance in this type of programme, and clarifies the kinds of additional data that would be needed to properly study social support and study groups. For researchers, it offers a cautionary tale about the limits of "network for network's sake"; for practitioners, it provides guidance on where to allocate scarce analytic and data collection resources.

# 6. LIMITATIONS AND FUTURE WORK

As with any single-institution, single-programme study, the present findings are subject to several limitations that delineate the scope of the conclusions and suggest avenues for further research.

## 6.1. Data scope and administrative constraints

First, the analysis relies exclusively on **administrative records** from a single Civil Engineering programme at a public university in Argentina. While this provides rich longitudinal coverage of enrolments, grades and institutional statuses, it does not include finer-grained digital traces (e.g., learning management system activity) or systematic survey data. As a result, the feature space is necessarily constrained to what the institution routinely records. Although this limitation reflects the reality of many universities, it restricts the types of social and behavioural constructs that can be operationalised.

Secondly, the construction of both curriculum-graph and network features depends on the **accuracy and completeness** of institutional data. While extensive cleaning and validation steps were undertaken, undetected errors or idiosyncratic recording practices may still affect specific variables. For example, variations in how course sections or exam sittings are coded can influence the structure of co-enrolment graphs (Gardner & Brooks, 2018; Lu et al., 2022). The robustness of the negative result regarding administrative SNA should therefore be revisited as similar pipelines are deployed in other institutions with different data systems.

**6.2. Measurement of social integration**

A central limitation is that the study uses **co-enrolment and co-exam co-occurrence** as proxies for social integration. As discussed in Section 5, sharing a course or an exam session is a very coarse indicator of potential interaction and may bear only a weak relationship to the kind of peer support, friendship ties or study groups emphasised in integration theories (Castro-Montoya & Ríos-Martínez, 2025; Tinto, 1993). Administrative co-occurrence networks may capture "synchrony of scheduling" more than genuine social embedding.

Future research should integrate **richer relational data** to better approximate the constructs of interest. Possible directions include:

- networks based on group projects, laboratory teams or formal mentoring relationships.

- interaction networks derived from discussion fora, collaborative tools or messaging in learning platforms.

- targeted surveys on perceived peer support, belonging and study group participation, linked to administrative records.

Such data would allow more direct tests of whether **genuinely relational networks** add predictive value beyond curriculum structure and performance, and whether specific types of ties (e.g., reciprocal friendships vs. weak academic contacts) play distinct roles in persistence (Ramsey et al., 2023; Siemens & Dawson, 2025; Woolcott & Lloyd, 2025).

**6.3. Generalisability beyond a single programme**

The study is conducted in a **single, strongly curriculum-constrained engineering programme**. The finding that administrative SNA does not improve dropout prediction may not generalise to programmes with:

- more flexible curricula and abundant electives.

- modular credit systems where students design highly individualised pathways.

- different class sizes or teaching formats (e.g., small seminars vs. large lectures).

In less constrained settings, administrative co-enrolment patterns might capture more meaningful variation in peer networks, simply because students have more freedom to cluster with peers or follow divergent pathways. Comparative studies across programmes with different levels of structural rigidity would help clarify how the **interaction between curriculum design and network formation** shapes the utility of administrative SNA.

Furthermore, the institutional and cultural context—public university, Argentina, high levels of economic volatility and periodic academic strikes—may influence both dropout dynamics and the role of peer support. Replication in other countries, institutional types and disciplines is necessary to establish the broader external validity of the results.

### 6.4. Modelling choices and alternative approaches

The modelling pipeline in this study uses **random forest classifiers** with group-wise feature addition to assess incremental value. Although this is a standard and robust approach, other algorithms and modelling strategies could be explored. Gradient boosting machines, calibration-sensitive methods, or survival models that explicitly model time-to-event might capture different aspects of risk (Fernberg, 2025; Khatun et al., 2025). Similarly, more advanced techniques for modelling interaction between structural and network features—such as graph neural networks or multi-layer graph embeddings—could potentially exploit patterns that tree-based ensembles treat as noise.

However, any future work along these lines should maintain a **strong emphasis on leakage control and interpretability**. As the initial leakage episode in this project illustrated, complex models trained on rich administrative datasets are particularly vulnerable to subtle violations of temporal ordering (Hartmann et al., 2022; Wang et al., 2022). Methodological innovations should therefore be accompanied by equally rigorous safeguards around feature engineering and validation schemes.

### 6.5. From prediction to intervention

Finally, even a highly accurate model does not by itself constitute an effective intervention strategy. The current study focuses on **predictive performance and feature importance**, not on the design or evaluation of concrete interventions. Future work should link predictive models to **actionable decision rules**, exploring, for example:

- how early in the programme reliable risk signals emerge.
- which combinations of structural and performance indicators best identify students for specific supports (e.g., targeted tutoring in bottleneck courses, alternative pacing through the curriculum);
- how to integrate qualitative insights from staff and students into model-driven early-warning systems.

Bridging the gap from prediction to intervention will require interdisciplinary collaboration between data scientists, programme directors, teaching staff and student support services, as well as careful ethical consideration of how predictive labels are communicated and used.

## 7. CONCLUSIONS

This study examined whether **administrative social networks**—constructed from co-enrolment and co-exam data—provide additional predictive value for student dropout in a **curriculum-constrained Civil Engineering programme**, once a leakage-aware early-performance model and curriculum-graph structural features are in place.

Using a three-semester observation window, a 16-fold leave-cohort-out design and a carefully audited CAPIRE data layer, we compared four models: a baseline CAPIRE configuration (M0), the same baseline plus administrative network features (M1), baseline plus curriculum-graph features (M2), and a full model combining all three (M3). After removing two post-outcome administrative variables that had produced artificially perfect performance, the **leakage-free results** were clear and stable across cohorts:

- The **baseline** and **baseline + graph** models (M0, M2) achieved **high and virtually identical performance** (F1 = 0.9411; ROC–AUC = 0.9776).

- Adding administrative network features **never improved prediction**. Both network-augmented models (M1, M3) yielded **lower F1-scores (0.9367)** and slightly lower ROC–AUC (0.9768), representing a small but consistent degradation of around **0.4 percentage points in F1**.

- Feature importance profiles showed that **early performance and curriculum-graph indicators dominate the predictive signal**, while co-enrolment-based network metrics occupy lower ranks and contribute little unique information.

Substantively, these findings suggest that in a **strongly structured engineering curriculum**, the main drivers of observable dropout risk are **structural and academic** rather than those aspects of social integration that can be inferred from administrative co-occurrence. Progression through bottlenecks, the accumulation of blocked credits and early success or failure in foundational courses shape both students' structural position in the curriculum and their synchrony with their cohort. Administrative co-enrolment networks largely reflect these structural dynamics rather than capturing an independent layer of peer support or study-group effects.

For **theory**, the study does not challenge the idea that social integration matters for persistence, but it shows that **co-enrolment alone is a poor operationalisation** of that construct in curriculum-constrained programmes. If the goal is to understand how genuine study groups, friendship ties or mentoring relationships protect against dropout, more **direct relational data**—from group work, interaction logs or surveys—will be needed.

For **institutional analytics**, the results point to two priorities:

1. Invest in **high-quality curriculum-graph modelling** and **early-performance monitoring** as the backbone of early-warning systems in structured programmes.

2. Be cautious about deploying complex administrative SNA pipelines unless there is evidence that the specific network data available align with theoretically meaningful constructs and add demonstrable predictive value.

Methodologically, the work underscores the importance of **data leakage audits** in learning analytics. The initial, implausible F1 = 1.00 result was only explainable by the presence of post-outcome variables; once these were removed and the pipeline rebuilt under the CAPIRE VOT framework, performance dropped to realistic levels and the true behaviour of graph and network features became visible. Reporting this process is itself a contribution: it illustrates how easily leakage can contaminate rich administrative datasets, and how a principled temporal architecture can help prevent it.

Finally, the study highlights the constructive role of **negative results**. The fact that administrative co-enrolment networks did *not* deliver the hoped-for gains is valuable information. It narrows the search space for effective predictive features, redirects analytic effort towards curriculum structure and early performance, and clarifies what kinds of additional data would be required to study the social mechanisms that theories of integration rightly emphasise. In that sense, the "failure" of administrative SNA in this context becomes a concrete guide for future work on dropout in complex, structurally constrained programmes.

**REFERENCES**

Bayer, J., Bydžovská, H., Géryk, J., Obsivac, T., & Popelínský, L. (2012). Predicting drop-outs from social behaviour of students. In *Proceedings of the 5th International Conference on Educational Data Mining* (pp. 103–109). International Educational Data Mining Society.

Breiman, L. (2001). Random forests. *Machine Learning*, 45(1), 5–32.

Castro-Montoya, B., & Ríos-Martínez, L. (2025). A cultural adaptation of Tinto's student integration theory in Latin America. *Cogent Education*, 12(1), 2479384. https://doi.org/10.1080/2331186X.2025.2479384

Fahd, K., Ratnapala, I., & Yuan, R. (2023). Effectiveness of data augmentation to predict students at risk of academic failure using educational data mining. *Social Network Analysis and Mining*, 13, 42. https://doi.org/10.1007/s13278-023-01117-5

Fernberg, E. (2025). *Early detection of university drop outs using machine learning* (Master's thesis, Arcada University of Applied Sciences).

Gardner, J. P., & Brooks, C. (2018). Learn from your (Markov) neighbour: Co-enrollment networks and their relationship to grades in undergraduate education. *Journal of Learning Analytics*, 5(3), 30–48. https://doi.org/10.18608/jla.2018.53.3

Hansen, J., Silva, A., & Heileman, G. (2024). Extending curricular analytics to analyse undergraduate physics pathways. *Physical Review Physics Education Research*, 20(2), 020143. https://doi.org/10.1103/PhysRevPhysEducRes.20.020143

Hartmann, J., Lörch, M., & Löffler, F. (2022). Data leakage through click data in virtual learning environments. In *Proceedings of the 14th International Conference on Educational Data Mining* (pp. 705–709). International Educational Data Mining Society.

Heileman, G. L., Hickman, M., Slim, A., & Abdallah, C. T. (2018). Curricular analytics: A framework for quantifying the impact of curriculum design on student success. arXiv. https://arxiv.org/abs/1811.09676

Hollar, J. (2023). *Analyzing the effect of bottleneck courses on time to graduation* (Master's thesis, Appalachian State University).

Khatun, M. R., Hasan, M. M., & Rahman, M. M. (2025). A hybrid framework of statistical, machine learning, and deep learning approaches for analysing factors influencing student dropout. *PLOS ONE*, 20(1), e0331917. https://doi.org/10.1371/journal.pone.0331917

López-Zambrano, J., Lara, J. A., & Romero, C. (2021). Early prediction of student learning performance through data mining: A systematic review. *Psicothema*, 33(4), 564–575. https://doi.org/10.7334/psicothema2021.60

Lu, S., Liu, Q., & Zhou, T. (2022). Academic failures and co-location social networks in campus. *EPJ Data Science*, 11, 35. https://doi.org/10.1140/epjds/s13688-022-00322-0

McCubbin, I. (2003). *An examination of criticisms made of Tinto's 1975 student integration model of attrition* (Unpublished manuscript). University of Glasgow.

Molontay, R., & Hülber, L. (2020). Characterising curriculum prerequisite networks by a probabilistic student flow model. *Applied Network Science*, 5, 62. https://doi.org/10.1007/s41109-020-00311-2

Padhye, S. M., Brawner, C. E., & Camacho, M. M. (2024). Analysing trends in curricular complexity and extracting common curricular design patterns. In *Proceedings of the ASEE Annual Conference & Exposition*. American Society for Engineering Education.

Pather, S. (2019). A conceptual framework for understanding student integration and persistence in open distance learning. *International Review of Research in Open and Distributed Learning*, 20(2), 221–241.

Paz, H. R. (2025a). A leakage-aware data layer for student analytics: The CAPIRE framework for multilevel trajectory modeling. arXiv. https://doi.org/10.48550/arXiv.2511.11866

Paz, H. R. (2025b). The CAPIRE Curriculum Graph: Structural feature engineering for curriculum-constrained student modelling in higher education. arXiv. https://doi.org/10.48550/arXiv.2511.15536

Ramsey, L. R., Etienne, B., & Smith, J. (2023). A social network analysis of STEM students at a regional university. *Journal of College Student Retention: Research, Theory & Practice*. Advance online publication. https://doi.org/10.1177/15210251231215787

Schellpfeffer, S. (2024). *An analysis of curricular complexity patterns in engineering majors* (Doctoral dissertation, University of North Dakota).

Siemens, G., & Dawson, S. (2025). Exploring social network analysis in education. *ACM Transactions on Computing Education*, 25(1), 1–28. https://doi.org/10.1145/3707463

Slim, A., Heileman, G. L., Hickman, M., & Abdallah, C. T. (2025). Bridging structure, pass rates, and student outcomes through curricular analytics. In *Proceedings of the 18th International Conference on Educational Data Mining* (pp. 255–260). International Educational Data Mining Society.

Stavrinides, P., Karamolegkos, P., & Voyiatzis, I. (2023). Course-prerequisite networks for analysing and improving curricula. *Applied Network Science*, 8, 29. https://doi.org/10.1007/s41109-023-00543-w

The Predictive Learning Analytics for Student Dropout Using Data Mining Technique: A Systematic Literature Review. (2025). *International Journal of Emerging Technologies in Learning*, 20(3), 45–62.

Tinto, V. (1993). *Leaving college: Rethinking the causes and cures of student attrition* (2nd ed.). University of Chicago Press.

Vaarma, M., Korhonen, V., & Laakso, M.-J. (2024). Predicting student dropouts with machine learning: A case study in higher education. *Studies in Higher Education*, 49(6), 1123–1142. https://doi.org/10.1016/j.stueduc.2024.101322

Wang, Y., Xie, P., & Yu, Z. (2022). Effectiveness of artificial intelligence models for predicting student dropout: A meta-analytic review. *Journal of Educational Computing Research*, 60(7), 1725–1755.

Woolcott, G., & Lloyd, M. (2025). Collaborative engagement versus collective estrangement: Social ecology networks in undergraduate mathematics. *Higher Education Analytics*, 4(2), 129–152.

Yang, B., Chen, J., & Zhao, L. (2025). Analysis of student progression through curricular networks: A graph-based framework. *Electronics*, 14(15), 3016. https://doi.org/10.3390/electronics14153016

Zomer, R. (2007). *An examination of the applicability of Tinto's model of student persistence in Christian higher education* (Master's thesis, Taylor University).